\documentclass[aps,pra,groupedaddress,notitlepage,twocolumn, 10pt,longbibliography]{revtex4-1}
\usepackage{fullpage}
\usepackage{amsmath,amsfonts,amssymb}
\usepackage{braket}
\usepackage{url}
\usepackage{hyperref}
\usepackage{graphicx}
\graphicspath{{figures/}}

\newcommand{\addsutd}{Science, Mathematics and Technology Cluster,\\ Singapore University
of Technology and Design, 8 Somapah Road, 487372 Singapore}

\begin{document}

\title{Interference‑Protected Subradiance and Bound States in Nested Atomic Arrays}
\author{Bella Santosa}
\affiliation{\addsutd}
\author{Daniel Leykam}
\email{daniel_leykam@sutd.edu.sg}

\affiliation{\addsutd}

\begin{abstract}
Collective subradiant states in waveguide QED are highly sensitive to disorder, limiting their scalability and robustness. We propose a deterministic approach to engineering atom arrays based on a Minkowski sum construction, generating quasi‑disordered structures with built‑in correlations. This leads to mode-selective radiative coupling: interactions between dark modes are parametrically suppressed, while bright modes can hybridize. We study the stability of these subradiant and bound‑state‑like modes against moderate positional disorder. Our work provides a route to robust, analytically controllable subradiance through engineered quasi‑disorder, with direct relevance to atom–waveguide and circuit‑QED experiments.
\end{abstract}

\maketitle
\thispagestyle{empty}

\section{Introduction}

Collective light-matter interactions in atomic arrays have been widely explored as a platform for engineering and controlling novel photonic states \cite{RevModPhys.95.015002}. When an array of atoms is coupled to a one-dimensional (1D) waveguide, their emission and absorption properties are fundamentally affected by long-range photon-mediated interactions, giving rise to cooperative phenomena such as superradiance and subradiance~\cite{Albrecht_2019}, arising due to constructive and destructive interference amongst the emitters, respectively. Subradiance involves collective modes with suppressed decay rates, which are attractive for applications in quantum memory and coherent light storage \cite{RevModPhys.95.015002}.

Subradiance in ordered atomic arrays has been extensively explored for periodic atomic arrays coupled to waveguides \cite{RevModPhys.95.015002,PhysRevLett.125.253601}. In this ordered configuration, the most subradiant eigenstates in these systems exhibit a characteristic decay rate that scales by a factor of $\xi^2/N^3$ where $N$ is the number of atoms in the array and $\xi$ labels the eigenstates by increasing decay rates \cite{PhysRevLett.122.203605,PhysRevLett.125.253601}. These collective modes can be described by generalized Bloch waves. The single-excitation eigenstate structure is dependent on the spacing of the periodic array. At the Bragg condition where lattice spacing is an integer multiple of half the transition wavelength $\lambda_0/2$, coherent dipole-dipole interactions vanish and the system is reduced to the Dicke model with purely dissipative coupling~\cite{Albrecht_2019}. Away from the Bragg condition, both coherent and dissipative interactions compete, resulting to a multimode structure with propagating polaritons. Both of these regimes have been thoroughly characterized and shown to have long-lived localised ``dark" states \cite{RevModPhys.95.015002,Albrecht_2019,PhysRevLett.122.203605,PhysRevLett.125.253601,PhysRevA.100.063832}.

The role of disordered array on the subradiance phenomenon, on the other hand, has also been explored recently. Gjonbalaj et al. \cite{PhysRevA.109.013720} found distinct regimes of eigenstate localisation and resulting decay suppression. In the dilute regime where the average atomic spacing is greater than the transition wavelength ($a>\lambda_0$), disorder induces an analog of Anderson localization where eigenstates become localised within the bulk, resulting in halted energy transport to the radiative boundaries. Conversely, disorder in the dense regime ($a<\lambda_0$) creates dimerised atoms that form two-body singlet states analogous to Dicke model states. Suppression of decay in this regime is dependent on the disorder strength, where stronger disorder achieves better suppression more reliably by producing sufficiently close atomic dimers for dark state formation~\cite{PhysRevA.109.013720}. However, since the process of atom position determination is randomized, the lack of deterministic control prevents reliable engineering of subradiant properties. 

Other means of controlling interference to engineer strongly subradiant states include spatial modulation of the atom-waveguide coupling strength~\cite{2256-6x36}, changing the properties of the photonic waveguide modes, such as using quasiperiodic modulation to tune between localized, propagating, and critical spectra~\cite{fibonacciwaveguidequantumelectrodynamics}, tuning the emitter frequency with respect to the photonic band gap to control the interplay between long range and near-field coupling~\cite{PhysRevA.111.023707}, and the use of ``giant atoms'' in which atomic emitters are coupled to multiple positions along the waveguide~\cite{Kannan2020,10.1007/978-981-15-5191-8_12,PhysRevA.108.043709}. Experimentally, these kinds of waveguide QED models can be realized using superconducting circuits, where qubits play the role of artificial atoms~\cite{PhysRevX.12.031036, Zanner2022}, as well as photonic crystal waveguides coupled to optically trapped atom arrays~\cite{Kim2019,https://doi.org/10.1002/qute.202000008}.

In pursuit of an analytically tractable and engineerable waveguide system with improved decay suppression, we explore an alternative approach inspired by recent advances in computational material design. Park et al. demonstrated that hypergraph modelling can be applied to material design to drastically speed-up large multiparticle simulations by exploiting Minkowski sum constructions \cite{park2025hypergraphmodellingwavescattering}. This construction allows for the rapid generation of large multiparticle systems $\mathbf{R}$ using small seed subsystems $\mathbf{Q}_1$ and $\mathbf{Q}_2$ through set operations such as $\mathbf{R}=\mathbf{Q}_1 \oplus \mathbf{Q}_2$, enabling the optimization of simulations of $10^4$ particles using only 200 seed particles. This drastic reduction of number of design parameters allows for scattering simulations to be of time complexity $O(N^{1/2})$ rather than $O(N)$ \cite{park2025hypergraphmodellingwavescattering}. The build-in correlations between atom positions in this construction also results in many terms becoming identical, allowing for more their more efficient computation, which in this paper manifests as the obtaining of analytical solutions through a smaller Hamiltonian matrix. Taking inspiration from this implementation of hypergraph modelling, we adapt the Minkowski sum construction as means to to achieve an analytically tractable quasi-disordered array. We show in this paper that the resulting Hamiltonian of an atomic array produced by this construction possesses symmetries leading to the formation of long-lived subradiant modes, and study the fate of these states under disorder in the atom positions.

The remainder of this paper will be outlined as the following: In Section \ref{section: general theory} we will present the Minkowski sum construction and the useful symmetries in the matrix derivations more in depth. We also present a general calculation of an array produced by nesting an arbitrary array in a dimer array. In Section \ref{section: dimer arrays} we present the simplest case of dimer arrays nesting ($N_A=N_B=2$). In Section \ref{section: periodic dimer arrays} we present and analyze the results of expanding this system by nesting a periodic array in a dimer array to simulate the dark states arising from dimers in disordered arrays \cite{PhysRevA.109.013720}. Finally, in Section \ref{section: deeper nesting} we discuss further applications of the Minkowski sum construction by assessing the eigenstate structures of arrays resulting from deeper nestings ($N=N_AN_BN_C...$). We conclude with Sec.~\ref{section: conc}.

\section{Model}
\label{section: general theory}

We consider the single-excitation case of $N$ two-level atoms coupled to a 1D waveguide. Following previous works, we adopt the Markovian approximation \cite{RevModPhys.95.015002}, which assumes that the photon degrees of freedom relax on a much shorter timescale than atomic dynamics, thus allowing them to be traced out \cite{RevModPhys.95.015002}. The resulting effective Hamiltonian takes the form $\hat{H} = \hat{H}_0 + \hat{H}_{\mathrm{eff}}$~\cite{RevModPhys.95.015002,Albrecht_2019}, where:
\begin{equation}
    \hat{H}_0 = \omega_0 \sum_{n=1}^N \hat{\sigma}_n^\dagger\hat{\sigma}_n,
\end{equation}
is the Hamiltonian of the bare atoms and,
\begin{align}
    \hat{H}_{\mathrm{eff}} = - i\gamma_{1D}\sum_{n,n^{\prime}=1}^N e^{i k_z |x_{n}-x_{n^{\prime}}|} \hat{\sigma}_n^\dagger\hat{\sigma}_{n^{\prime}},
\end{align}
describes atom-atom interactions mediated by the waveguide. Here $\omega_0$ is the atoms' transition frequency, $\hat{\sigma}_n^\dagger$ and $\hat{\sigma}_n$ are the raising and lowering operators of atom $n$ respectively, $x_n$ is the position of the $n$-th atom, and $\gamma_{1D}$ is the single atom emission rate into the waveguide. The exponential factor encodes the phase accumulated by a photon propagating between atoms $n^{\prime}$ and $n$, giving rise to coherent and dissipative photon-mediated interactions between atoms. Interactions can generate subradiant states - eigenvectors of $\hat{H}$ with eigenvalues $\omega_{\xi}$ satisfying $\mathrm{Im}(\omega_\xi)\ll \gamma_{1D}$.

\begin{figure}[]
\centering
\includegraphics[width=\columnwidth]{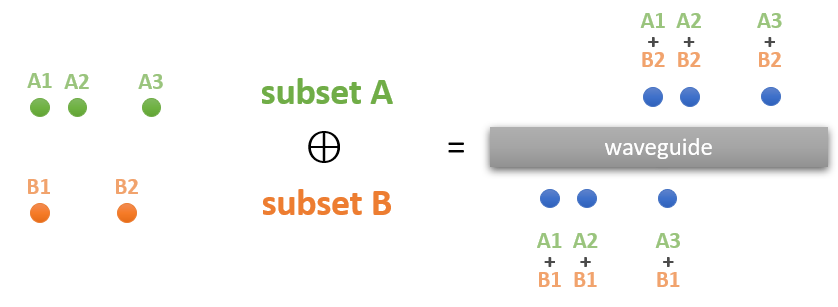}
\caption{Atom positions along one-dimensional array given by the Minkowski sum of an arbitrary subset array $\mathbf{X}_A$ and a dimer subset array $\mathbf{X}_B$ produces two copies of $\mathbf{X}_A$ offset by $d_B$. For the rest of this paper, shifted ``duplicates'' are presented as copies displaced vertically.}
\label{fig: Minkowski sum}
\end{figure}

Previous works have considered periodic atom arrays with uniform spacing $x_n = n d$, for which analytical solutions in terms of Bloch waves can be obtained, as well as disordered arrays in which the uniform spacings are randomly perturbed. In the latter, the longest-lived states are hosted by the atom pairs with the smallest separations~\cite{PhysRevA.109.013720}.

In the following, we will consider structured arrays generated from simpler seed systems. We define a set of atom positions as $\mathbf{X}^{(i)} = \{x_{1}^{(i)},...,x_{N_i}^{(i)}\}$, where $x_{n}^{(i)}$ is the position of the $n$th atom. Given two sets of positions $\mathbf{X}^{(A)}$, $\mathbf{X}^{(B)}$, we can construct a larger array from the Minkowski sum of the two subsets as,
\begin{align}
    \mathbf{X}^{(A \oplus B)} &= \{x_{n}^{(A)} + x_{m}^{(B)} \; | \; x_{n}^{(A)} \in\mathbf{X}^{(A)}, \; x_{m}^{(B)}\in\mathbf{X}^{(B)}\} \nonumber \\
    &= \{ x_{n,m} \}, 
\end{align}
where subscripts separated by commas are used to index the different seed sets. Because the sum is over all combinations of positions drawn from $\mathbf{X}^{(A)}$ and $\mathbf{X}^{(B)}$, the total number of atoms in $\mathbf{X}^{(A\oplus B)}$ is $N=N_A N_B$, as shown in Fig.~\ref{fig: Minkowski sum}. The effective Hamiltonian describing the resulting atom array is
\begin{equation}
\hat{H}_{\mathrm{eff}} = -i \gamma_{1D} \sum_{m,m^{\prime}=1}^{N_B} \sum_{n,n^{\prime}=1}^{N_A} e^{i k_z |x_{n,m} - x_{n^{\prime},m^{\prime}}|} \hat{\sigma}_{n,m}^\dagger\hat{\sigma}_{n^{\prime},m^{\prime}}.    
\end{equation}
Note that while two indices ($n$ and $m$) are used, the positions $x_{n,m}$ do not refer to a 2-dimensional (2D) array. The atom array resulting from the Minkowski sum is in general, not periodic even if $\textbf{X}^{(A)}$ and $\textbf{X}^{(B)}$ are both periodic. Rather, the Hamiltonian describes obtain $N_B$ copies of $\textbf{X}^{(A)}$ translated by distances drawn from $\textbf{X}^{(B)}$.

Because the effective interaction between atoms only depends on their absolute separation, the Minkowski sum construction grants $\hat{H}_{\mathrm{eff}}$ an additional block structure, 
\begin{widetext}
\begin{align}
    \hat{H}_{\mathrm{eff}}  
    &= -i\gamma_{1D} \sum_{m=1}^{N_B} \overbrace{\left( \sum_{n,n^{\prime}=1}^{N_A} e^{i k_z |x^{(A)}_{n}-x^{(A)}_{n^{\prime}}|} \hat{\sigma}_{n,m}^\dagger\hat{\sigma}_{n^{\prime},m} \right)}^{\hat{H}^{(A)}} - i\gamma_{1D} \sum_{n=1}^{N_A} \overbrace{\left( \sum_{m,m^{\prime}=1}^{N_B} e^{i k_z |x^{(B)}_{m}-x^{(B)}_{m^{\prime}}|} \hat{\sigma}_{n,m}^\dagger\hat{\sigma}_{n,m^{\prime}} \right)}^{\hat{H}^{(B)}} \nonumber \\
    & \qquad \qquad -i \gamma_{1D} \sum_{m \neq m^{\prime}} \sum_{n \neq n^{\prime}} e^{i k_z |x_{n,m} - x_{n^{\prime},m^{\prime}}|} \hat{\sigma}_{n,m}^\dagger\hat{\sigma}_{n^{\prime},m^{\prime}}.  \label{eq:heff_block} 
\end{align}
The first term in brackets in Eq.~\eqref{eq:heff_block} corresponds to blocks which are diagonal in the $\mathbf{X}^{(B)}$ space. Each block is identical because the internal spacings within each copy of $\mathbf{X}^{(A)}$ are invariant under translations of all its atoms by $x^{(B)}_m$. A similar symmetry holds for the second term in brackets in Eq.~\eqref{eq:heff_block}, corresponding to copies of $\mathbf{X}^{(B)}$ which are block diagonal in the $\mathbf{X}^{(A)}$ space. Thanks to this structure, the eigenstates of arrays with atom positions $\mathbf{X}^{(A)}$ (or, alternatively $\mathbf{X}^{(B)}$) will serve as a useful basis for understanding the response of the nested array.

Let us denote the right and left eigenstates of $\hat{H}_{\mathrm{eff}}^{(A)}$ with energy $\omega_{\mu}$ as:
\begin{align}
    \ket{\psi_{\mu,m}} = \sum_{n=1}^{N_A} \psi_{\mu n} \hat{\sigma}_{n,m}^{\dagger} \ket{0}, \quad
    \bra{\psi_{\mu,m}} = \sum_{n=1}^{N_A} \bra{0} \hat{\sigma}_{n,m} \psi_{\mu n}, \quad
    \braket{\psi_{\nu m } | \psi_{\mu  m}} = \sum_{n=1}^{N_A} \psi_{\nu n} \psi_{\mu n} = \delta_{\mu \nu},
\end{align}
where we use Greek subscripts to index modes and Latin subscripts to index the atoms. Note the the lack of complex conjugation in the expression for the left eigenstates, since $\hat{H}_{\mathrm{eff}}$ is a complex symmetric non-Hermitian matrix~\cite{moiseyev2011non}. In the basis of eigenstates of $\hat{H}_{\mathrm{eff}}^{(A)}$, the matrix elements of $\hat{H}_{\mathrm{eff}}$ are
\begin{equation}
    \braket{ \psi_{\nu,m^{\prime}}|\hat{H}_{\mathrm{eff}} | \psi_{\mu,m} }=  \omega_{\nu} \delta_{\nu \mu} \delta_{m m^{\prime}} -i  \gamma_{1D} \left( \delta_{\nu \mu} e^{i k_z |x^{(B)}_m - x^{(B)}_{m^{\prime}}|}  + \sum_{n\neq n^{\prime}}^{N_A} \psi_{\nu n^{\prime}} e^{i k_z |x_{n,m} - x_{n^{\prime},m^{\prime}}|} \psi_{\mu n} \right)(1-\delta_{m m^{\prime}}). \label{eq:eigenbasis}
\end{equation}
\end{widetext}
In this picture, the spectrum of a nested array can be understood in terms of couplings between copies of the modes of $\hat{H}_{\mathrm{eff}}^{(A)}$. The first term in brackets on the right hand side of Eq.~\eqref{eq:eigenbasis} resembles coupling in an ordinary 1D atom array. The second term can either increase or decrease this effective coupling, depending on the mode's phase profile and the relative positions of the atoms. 

\section{Nested dimer arrays}
\label{section: dimer arrays}

In this section, we consider the simplest application of the Minkowski sum construction: the nesting of a dimer subset array in another. Setting $N_A=N_B=2$, we obtain two dimer seed arrays with arbitrary spacings: $\mathbf{X}^{(A)}=\{0,d_A\}$ and $\mathbf{X}^{(B)}=\{0,d_B\}$. Applying the Minkowski sum construction yields the four-atom array:
\begin{equation}
    \mathbf{X}^{(A\oplus B)} = \{0,d_A,d_B,d_A+d_B\}.
\end{equation}
The effective Hamiltonian of this system can be written in the block form,
\begin{equation}
    \hat{H}_\text{eff} = \left( \begin{array}{cc} \hat{H}^{(A)}_\text{eff} & \hat{H}^{(AB)} \\ \hat{H}^{(BA)} & \hat{H}^{(A)}_\text{eff} \end{array} \right).
\end{equation}
To diagonalise, we begin with the eigensystem of $\hat{H}^{(A)}$ \cite{RevModPhys.95.015002}: $\omega_{\pm} = \omega_0 - i \gamma_{1D} (1 \pm e^{i \varphi})$ with eigenvectors $\ket{\psi_{\pm, m}}= (1, \pm 1)^T/\sqrt{2}$. We assume $k_z d_A < \pi/2$, such that the symmetric eigenvector $(1,1)^T$ is superradiant and the antisymmetric eigenvector $(1,-1)^T$ is dark. Given the known eigenvectors, the interaction term between the $\mathbf{X}^{(B)}$ copies $\hat{H}^{(AB)}$ can be evaluated analytically. The relative values of $d_{A,B}$ determine the form of the interaction term (based on how the absolute value simplifies). Without loss of generality, we can assume $d_B > d_A > 0$,
\begin{equation}
    \hat{H}^{(AB)} = i \gamma_{1D} e^{i k_z d_B} \left( \begin{array}{cc} 1 & e^{i k_z d_A} \\ e^{-i k_z d_A} & 1 \end{array} \right).
\end{equation}
Evaluating in the eigenbasis of $\hat{H}_A$:
\begin{align}
    \hat{H}^{(AB)} &= i \gamma_{1D} e^{i k_z d_B} \left( \begin{array}{cc} 2  \cos^2 (\frac{ k_z d_A}{2}) & i \sin (k_z d_A) \\ -i \sin (k_z d_A) &  2 \sin^2 (\frac{k_z d_A}{2}) \end{array} \right) \\ &\approx i \gamma_{1D} e^{i k_z d_B} \left( \begin{array}{cc} 2 - \frac{k_z^2 d_A^2}{2} & i k_z d_A \\ -i k_z d_A &  \frac{k_z^2 d_A^2}{2} \end{array} \right),
\end{align}
where the approximation is for small $k_z d_A \ll 1$. The interaction between the two arrays involves a phase shift tunable via their relative spacing $d_B$, controlling whether energy splittings occur in the real or imaginary terms of the eigenenergy. There is a strong interaction between the superradiant modes, whereas destructive interference strongly suppresses interactions involving the subradiant modes. This can also serve as a model for disordered arrays, i.e. interactions between pairs of atoms with the smallest spacings.

\begin{figure}[]
\centering
\includegraphics[width=\columnwidth]{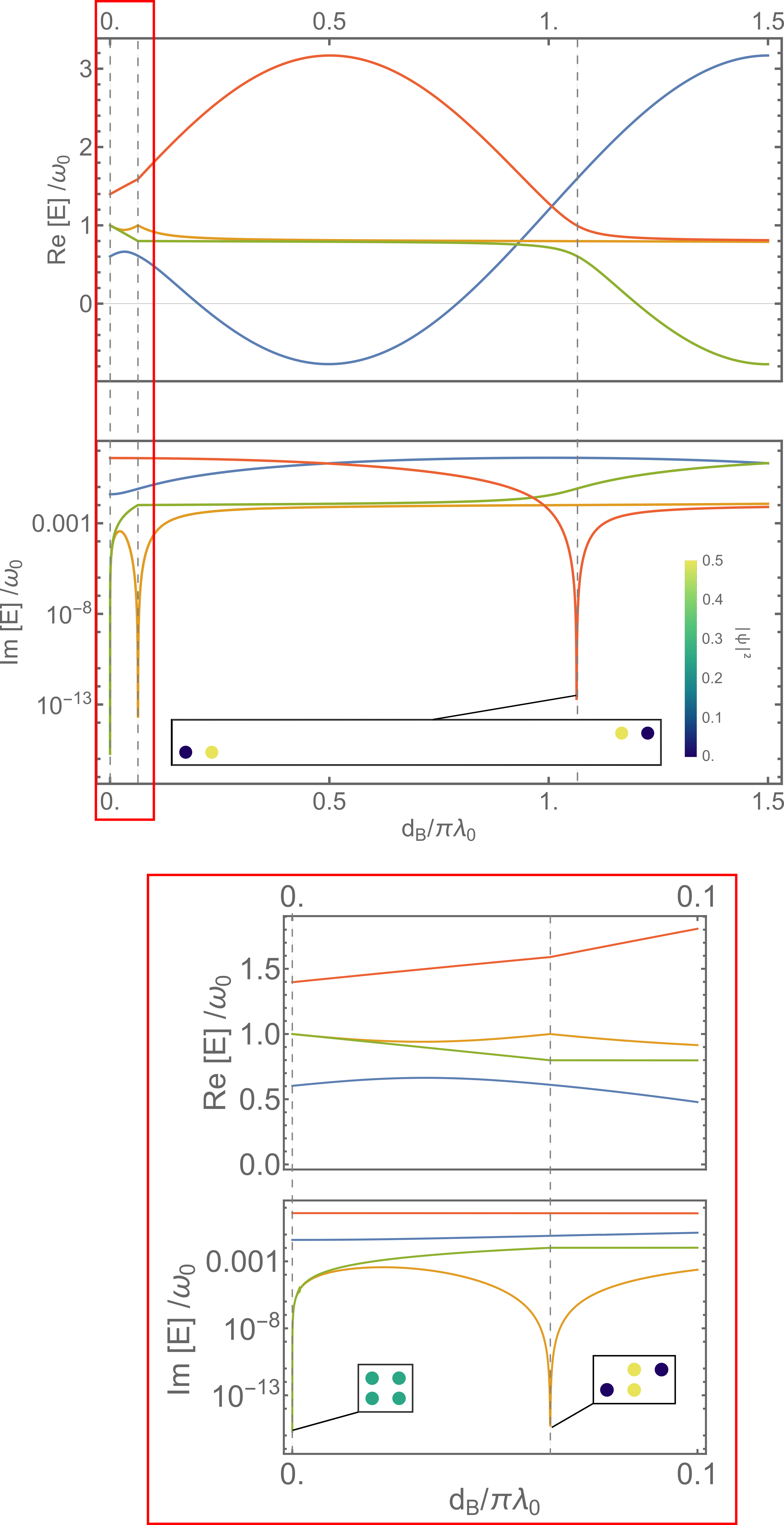}
\caption{Real and imaginary parts of the eigenvalues of a nested dimer system with $d_A=0.2\pi\lambda_0$ while varying $d_B/\lambda_0 \in [0,1.5\pi]$. We define two regimes: $d_A>d_B$ as regime I where the cluster copies are ``overlapping" (lower panels) and $d_A<d_B$ as regime II where the cluster copies are separated by a finite distance (upper panels). Insets show intensity profiles of eigenmodes for different resonant spacings $d_B/\lambda_0=\{0,0.2,1.065\pi\}$.}
\label{fig: nested dimers eigenstructure}
\end{figure}

Figure~\ref{fig: nested dimers eigenstructure} shows the energy spectrum of the array as a function of the spacing $d_B$. For $d_B < d_A$ (regime I), the two arrays partially overlap, resulting in sharp long-lived resonances (dark states) whenever a pair of atoms has the displacement. When $d_B=0$, there are two pairs of overlapping atoms resulting in two perfectly dark states and equal participation of all atoms in all modes. A single atomic position overlap also occurs at $d_A=d_B$ where another long-lived state occurs.

When $d_B > d_A$ (regime II), the two arrays are separated by a finite distance. In this case, the spectrum is dictated by the long-range interactions between the radiant modes, while the dark modes interact very weakly and have eigenvalues insensitive to $d_B$. At $d_B=1.065\pi\lambda_0$ an avoided crossing with the dark states leads to the formation of an extremely long-lived resonance. This subradiant state repeats at periodic intervals of $\Delta d_B=\pi\lambda_0$ in regime II and can be understood as a kind of bound state in the continuum (BIC)~\cite{kang2023applications} protected by the the symmetry between the two copies of $\mathbf{X}^{(A)}$.

\begin{figure}[]
\centering
\includegraphics[width=\columnwidth]{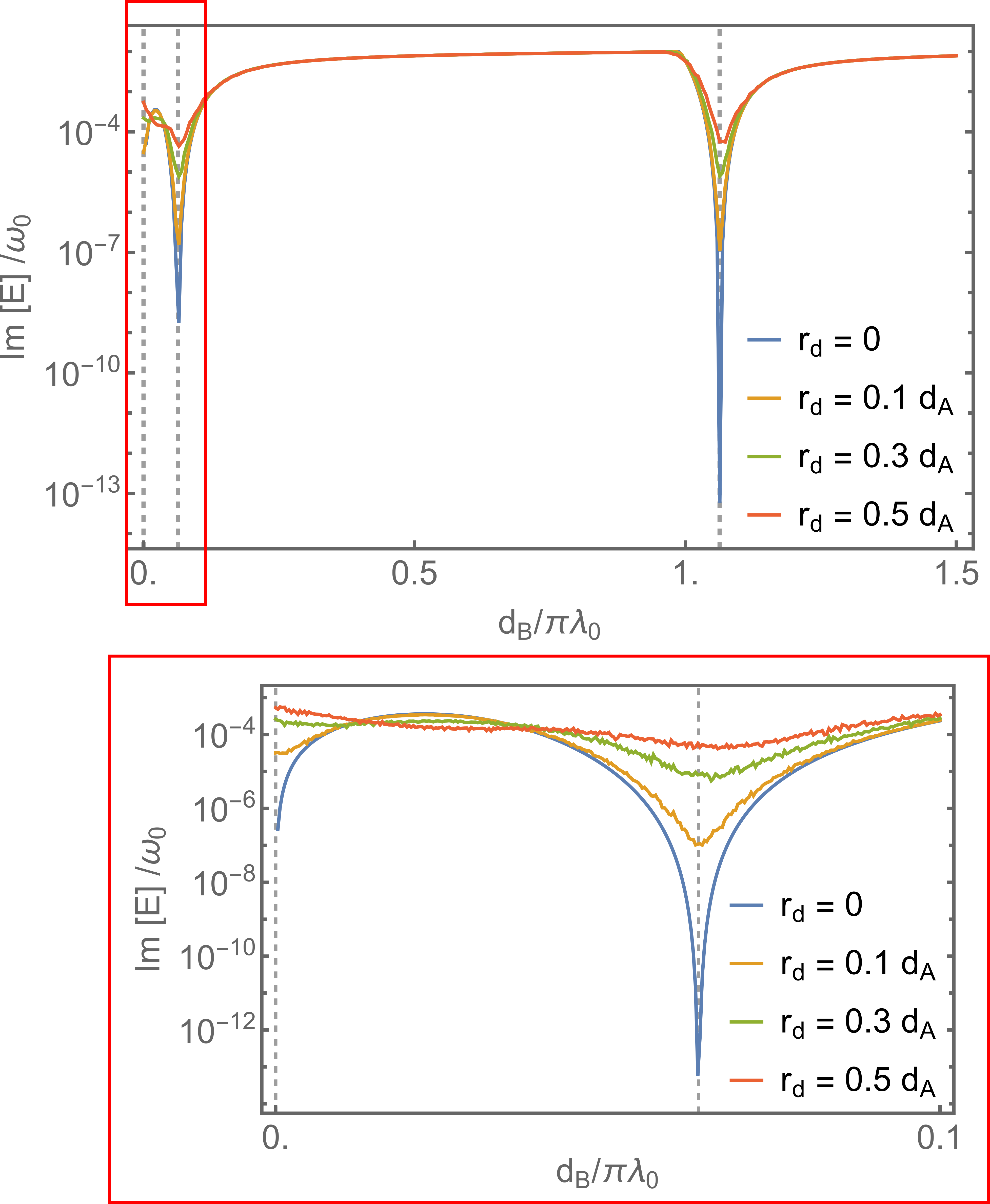}
\caption{Average imaginary part of the eigenvalue of the mode with the slowest decay rate in a nested dimer system while introducing various random disorder. Intra-dimer spacing is set as $d_A=0.2\pi\lambda_0$ and the separation of dimer copies are varied $d_B/\lambda_0 \in [0,1.5\pi]$. Random disorder of an arbitrary deviation of atom positions within the range $[-r_d,r_d]$ are introduced and averaged over $200$ samples. The ideal case of no disorder ($r_d=0$) is also plotted for comparison. At notable points ($d_B=d_A=0.2\pi\lambda_0$ and $d_B=1.065\pi\lambda_0$), the subradiant mode resonances remain pronounced for sufficiently small disorder in the atom positions, $r_d=0.1d_A$.}
\label{fig: nested dimer disorder}
\end{figure}

In realistic arrays unavoidable disorder will break this symmetry, decreasing the lifetime of the dark states. We model disorder by randomly perturbing the atom positions as $x_{n,m} \rightarrow x_{n,m} + \epsilon_{n,m}  r_{d}$, where $\epsilon_{n,m} \in [-1,1]$ is a random variable and $r_d$ is the disorder strength. Fig.~\ref{fig: nested dimer disorder} shows the decay rate of the longest-lived mode averaged over 200 disorder realizations, for various disorder strengths. We see that disorder comparable to the smaller spacing $d_A$ already gives a significant reduction in the lifetime of the dark states.

\section{Dimer-nested periodic array}
\label{section: periodic dimer arrays}

While the nested dimer provides a minimal setting to understand interference in nested arrays constructed by the Minkowski sum, its small size limits the number and disorder robustness of the resulting dark modes. Larger periodic subwavelength-spaced arrays exhibit a whole family of dark states with differing lifetimes~\cite{RevModPhys.95.015002,Albrecht_2019,PhysRevLett.122.203605,PhysRevLett.125.253601,PhysRevA.100.063832}. In this Section we study spectrum obtained by nesting arrays of several atoms.

\begin{figure}[]
\centering
\includegraphics[width=\columnwidth]{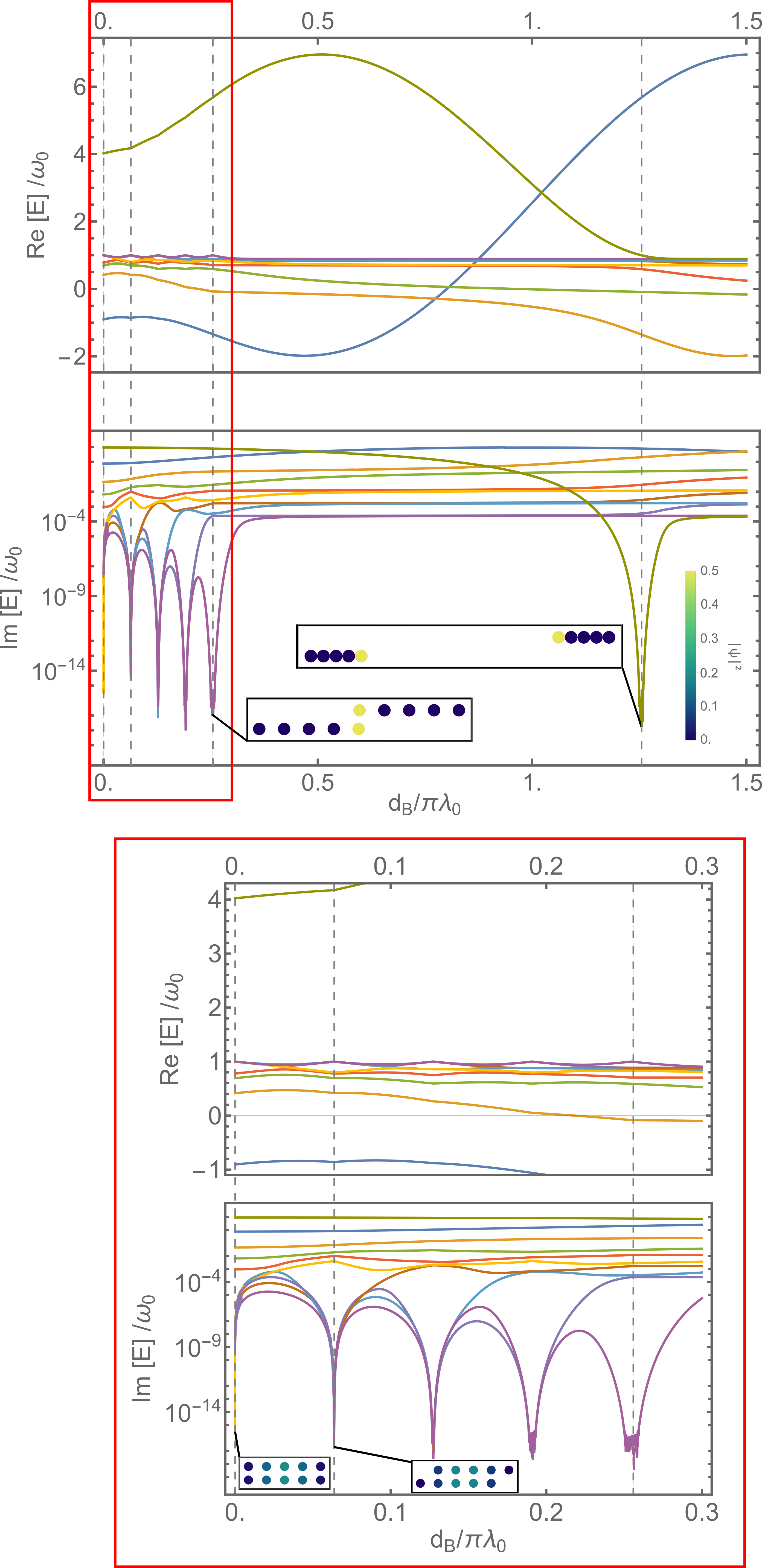}
\caption{Real and imaginary parts of the eigenvalues of a periodic array $\mathbf{X}^{(A)}=\{0,0.2,0.4,0.6,0.8\} \lambda_0$ nested in a dimer array $\mathbf{X}^{(B)}=\{0,d_B\}$ while varying $d_B/\lambda_0 \in [0,1.5\pi]$. Insets show intensity profiles of eigenmodes for $d_B/\lambda_0=\{0,0.2,0.8,1.256\pi\}$.} 
\label{fig: nested N5 eigenstructure}
\end{figure}

Fig.~\ref{fig: nested N5 eigenstructure} shows the spectrum of a $N_A = 5$ site subwavelength-spaced ($d_A = 0.2 
\lambda_0$) periodic array nested with a dimer with spacing $d_B$. We observe a pair of bright modes with short lifetimes along with a set of dark modes with much longer lifetimes. 

The eigenvalues of the bright modes are sensitive to $d_B$ and well-described by a perturbation theory that only takes into account the $\nu = \mu$ matrix elements in Eq.~\eqref{eq:eigenbasis}. In the limit of small spacing, the superradiant mode of the $\mathbf{X}^{(A)}$ array has a uniform amplitude~\cite{RevModPhys.95.015002}; constructive interference amongst the atomic dipoles enhances the effective coupling strength by a factor of $N_A$. Similar behaviour occurs for bright modes of nested arrays with larger $N_B$.

The number of subradiant modes present can be associated to the number of overlapping atoms in each corresponding atom configuration. Note that these subradiant modes in regime I (overlapping arrays) are insensitive to variations in $d_B$. In regime II (separated arrays), a single subradiant mode is present at periodic intervals of $d_B$ values (as shown in Fig.~\ref{fig: nested N5 eigenstructure} at $d_B=1.256\pi\lambda_0$) separated by $\Delta d_B=\pi\lambda_0$. Also similar to the nested dimer case, the subradiant mode in regime II can be associated with long-range interactions, as seen in the significant participation of the nearest atomic positions across the cluster duplicates.

\begin{figure}[]
\centering
\includegraphics[width=\columnwidth]{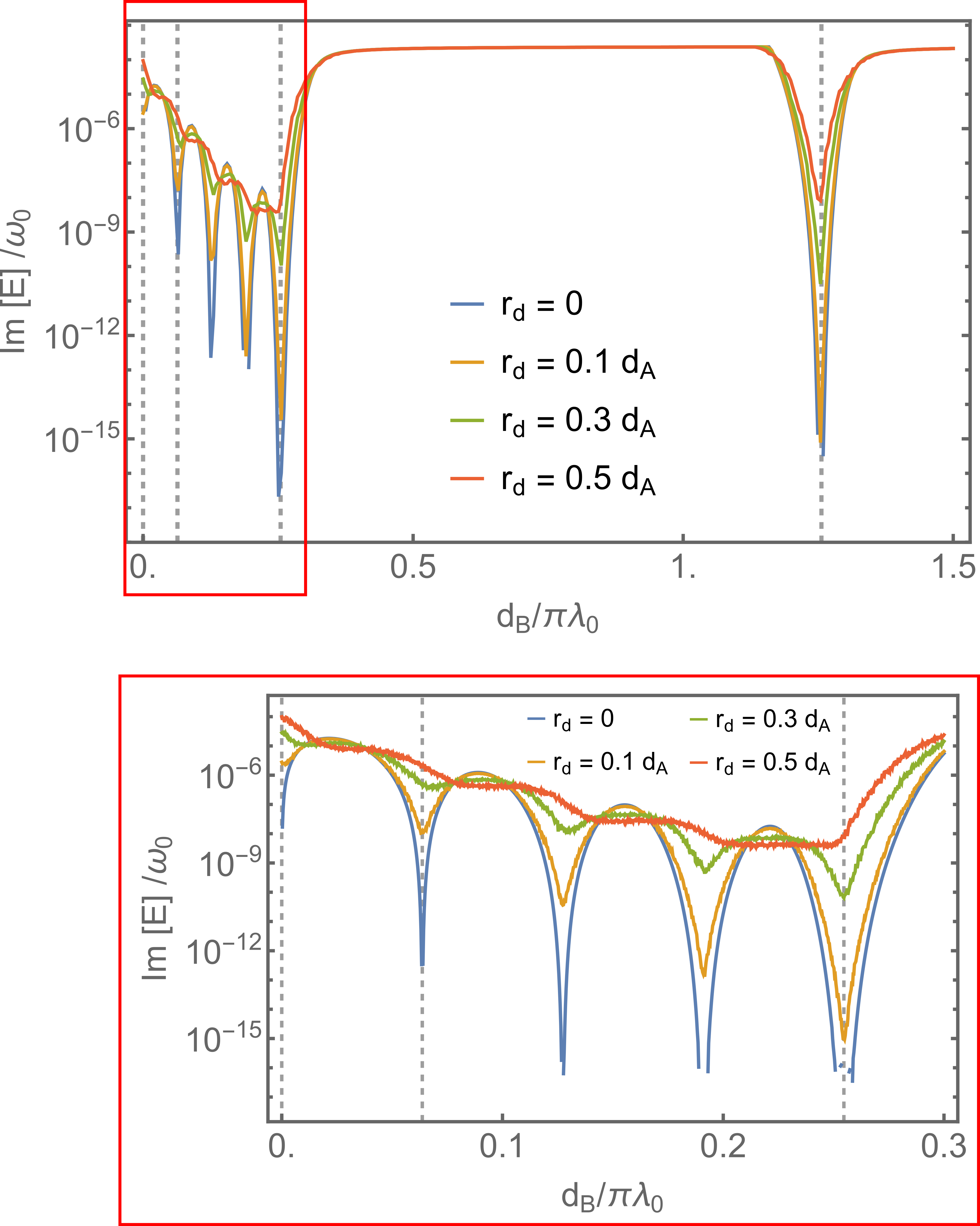}
\caption{Average imaginary part of the eigenvalue of the mode with the slowest decay rate in a periodic array $\mathbf{X}^{(A)}=\{0,0.2,0.4,0.6,0.8\} \lambda_0$ nested in a dimer array $\mathbf{X}^{(B)}=\{0,d_B\}$ while introducing random disorder $r_d$.}
\label{fig: nested N5 disorder}
\end{figure}

It is interesting to compare the two types of long-lived modes which can emerge in the nested array - a defect state, localized between the two arrays (at the edge of each) for separated copies, and families of defect modes that appear for partially overlapping arrays. In this case, the centre of the array forms a staggered array (with alternating spacings determined by $d_{A,B}$), embedded in a cladding array with a uniform spacing $d_A$. This cladding array acts as reflectors of the eigenmodes that fall outside of its passband, thus confining the subradiant modes at the centre of the array, as seen in the eigenvector profiles in the insets of Fig.~\ref{fig: nested N5 eigenstructure}.

The collective reflection provided by the cladding array leads to less sensitivity of the longest lifetime to disorder in the atom positions, as shown in Fig.~\ref{fig: nested N5 disorder}. In regime I where the two arrays are partially overlapping ($d_B<N_Ad_A$), the darkest modes are relatively insensitive to disorder in a significantly larger window of $d_B$ with increased size of nested array. We also observe the increased insensitivity against disorder with larger cladding arrays, demonstrating the significance of cladding in preserving these subradiant states. The subradiant mode in regime II is also similarly insensitive against moderate disorder, as seen in the preservation of a dark state with $\text{Im}[E]\approx10^{-8}\gamma_{1D}$ with the introduction of disorder $r_d=0.5d_A$.

\section{Deeper nested arrays}
\label{section: deeper nesting}

The emergence of sharp, interference‑protected subradiant modes in nested arrays naturally raises the question of how these effects evolve as the nesting depth increases. The Minkowski sum structure provides a simple way to construct such arrays using a small number of tunable parameters. Deeper nestings introduce an additional hierarchical structure in which multiple length scales, mode families, and symmetry sectors can interact,thus allowing for the engineering of further quasi-disorder while remaining analytically tractable by inheriting the eigenbasis derived from each previous nesting.

\begin{figure}[]
\centering
\includegraphics[width=\columnwidth]{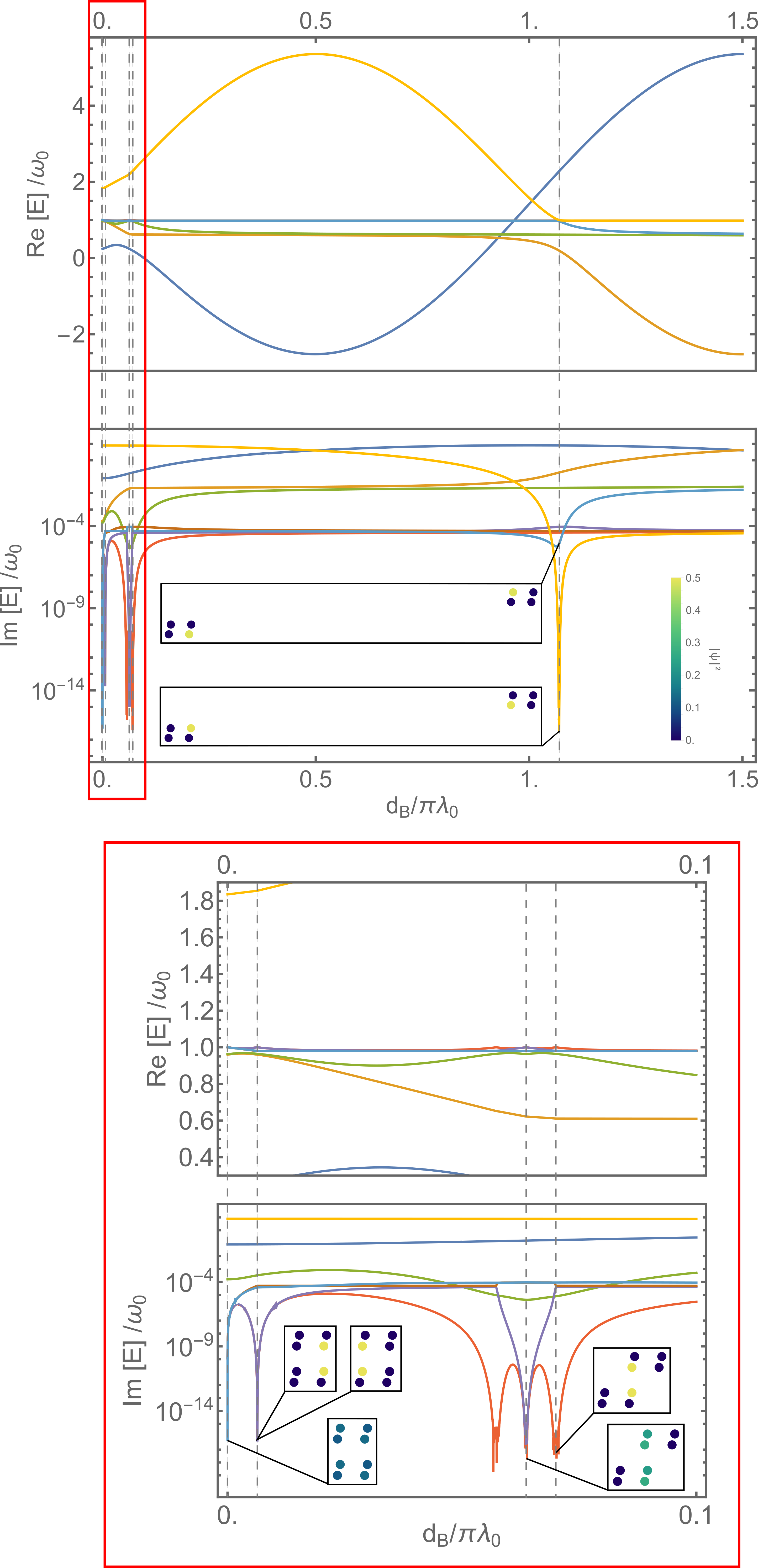}
\caption{Real and imaginary parts of the eigenvalues of doubly nested dimer arrays, fixing $d_A=0.2\pi\lambda_0$ and $d_C=0.02\pi\lambda_0$, as a function of $d_B/\lambda_0\in[0,1.5\pi]$. Insets show intensity profiles of eigenmodes for $d_B/\lambda_0=\{0,0.02,0.2,0.22,1.07\pi\}$ highlighting some notable points where subradiance is present. Each dimer copy in these participation plots are represented with perpendicular displacement to the direction of separation for clarity but do not represent variations in coupling strength of atoms to the waveguide.}
\label{fig: double nested eigenstructure}
\end{figure}

We demonstrate this inheritance of characteristics from a shallower nesting by introducing a deeper nesting to the single nested case presented in Sec.~\ref{section: dimer arrays}. Fig.~\ref{fig: double nested eigenstructure} illustrates the spectrum of the simplest case of double nesting, doubly nested dimers, described by three separations $d_A, d_B, d_C$. We keep $d_A$ and $d_B$ as in Sec.~\ref{section: dimer arrays}, while fixing a deeply subwavelength $d_C = 0.02\pi \lambda_0$.

The doubly nested case preserves characteristics of the singly nested dimers, where two sharp subradiant modes arise in regime I at atomic position overlaps. However, this additional nesting results in additional atomic positions to account for parameterised by $d_C$. Analogous to the first subradiant mode in regime I for the single nested dimers, two subradiant modes are present at $d_B=d_C$ corresponding to the two pairs of overlapping atomic positions. This configuration can be seen as two single nested dimers $d/\lambda_0=\{0.02,0.02\}$, separated by $d_A/\lambda_0=0.2$ where each of their dark states are localised in the central overlapping atom positions and are non-interacting across the nested dimers, as highlighted in the Fig.~\ref{fig: double nested eigenstructure} insets. 

\begin{figure}[]
\centering
\includegraphics[width=\columnwidth]{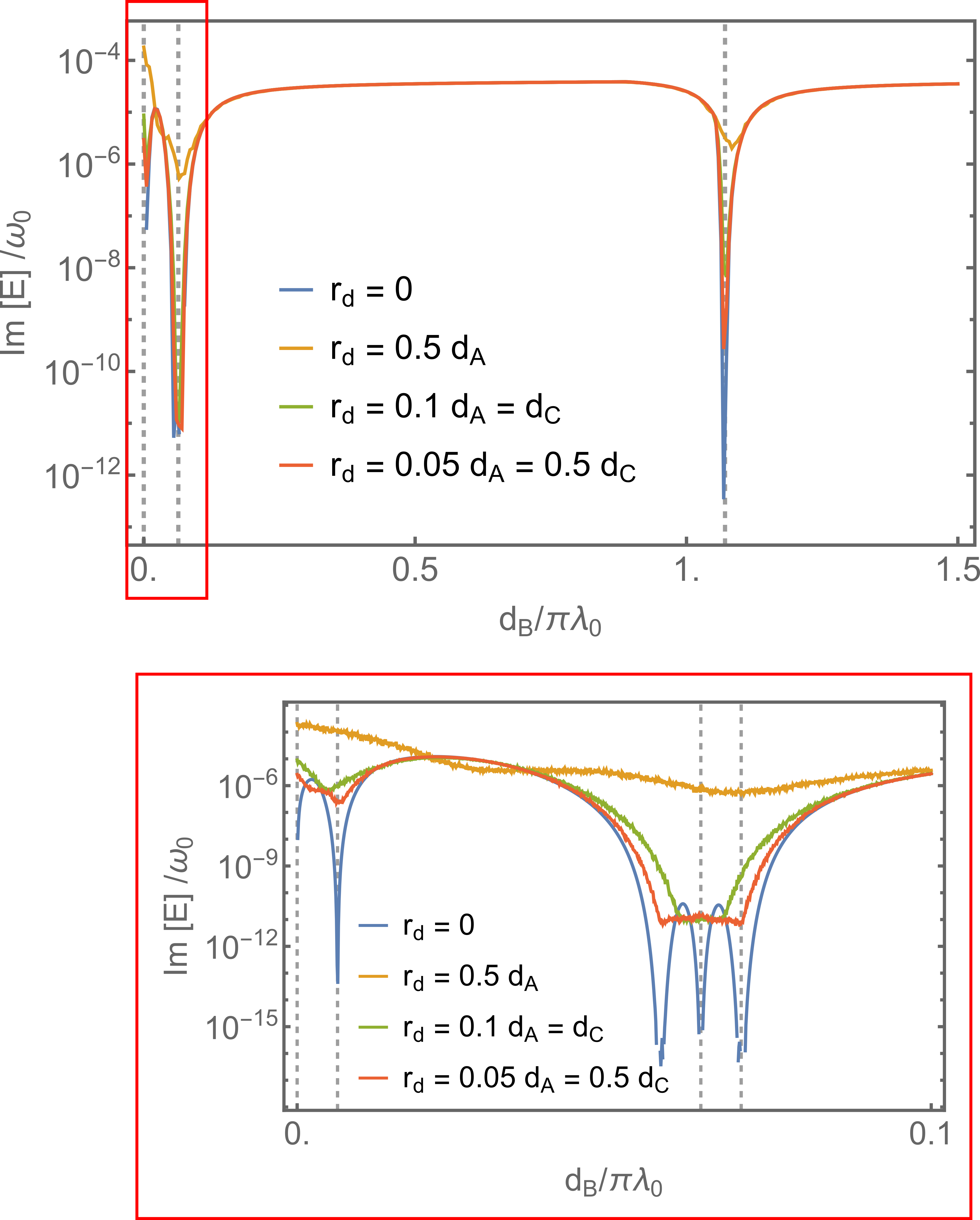}
\caption{Average imaginary part of the eigenvalue of the darkest mode in doubly nested dimer arrays with introduction of random disorder $r_d$.}
\label{fig: double nested disorder}
\end{figure}

In the subradiance seen previously in the single nested case at $d_A=d_B$, we observe a second subradiant mode corresponding to the overlap of the two central pairs of atom positions cladded by dimers on both sides. This second dark mode is also subradiant at $d_B=d_A\pm d_C$, presenting similarly to energy level ``splittings", which results in an extended resonance range of a relatively dark mode for $d_B\in[d_A-d_C,d_A+d_C]$. Within the extended window of darker modes, the subradiant states are therefore less sensitive to random disorders compared to a shallower nesting due to an increased density of subradiant peaks, as shown in Fig.~\ref{fig: double nested disorder}. This insensitivity may be reduced with increased cladding through deeper nesting of larger atomic arrays. 

Similarly to regime II in Sec.~\ref{section: dimer arrays}, a single subradiant mode is present at periodic intervals of $\Delta d_B=\pi\lambda_0$, visible in Fig.~\ref{fig: double nested eigenstructure} at $d_B=1.07\pi\lambda_0$. Unlike the single nested array, however, multiple bands of eigenmodes insensitive to $\Delta d_B$ are observed, which can be attributed to the increased number variations of length scales arising from a deeper nesting.

\section{Conclusion}
\label{section: conc}

In summary, the Minkowski sum construction provides a deterministic and semi-analytically transparent route to engineering subradiant states in waveguide QED arrays. By nesting simple seed structures (dimers and periodic chains), we revealed the emergence of interference-protected dark modes bound to interfaces between array copies. The resulting block-structured Hamiltonians expose clear mechanisms for suppressing radiative loss through built‑in positional correlations, enabling both long-lived defect states in overlapping regions and symmetry-protected bound-state-like modes at large separations. These features persist under moderate disorder, offering a flexible design principle for realizing stable subradiant excitations in current atom–waveguide and circuit-QED platforms.

It will be interesting to generalize the present approach to the creation of long-lived multiphoton states~\cite{PhysRevResearch.3.033233, PhysRevA.110.053707,PhysRevA.101.043845}. For example, one can envisage long-lived states in which each photon resides in a subradiant mode of a different copy of the initial Hamiltonian $\hat{H}^{(A)}$. 

The tensor product structure of the nested Hamiltonian also resembles the structure of hierarchically nested lattices recently introduced in the context of optical ring resonators~\cite{mehrabad2025quantummetamorphosisprogrammableemergence}. One important difference here is the long range coupling between the copies, as opposed to the limited local coupling considered in Ref.~\cite{mehrabad2025quantummetamorphosisprogrammableemergence}. Studying connections to higher order network theory and quantum quantum hypergraph states~\cite{BATTISTON20201,Leykam31122023,Rossi_2013} are other possible directions for future work.

\section*{Acknowledgements}

We acknowledge support from the Ministry of Education, Singapore, under its SUTD Kickstarter Initiative (Grant No. SKI 20210501).

\bibliography{references.bib}

\end{document}